\documentclass{article}
\usepackage[T1]{fontenc}
\usepackage{spconf,amsmath,graphicx,booktabs,multirow,hyperref}
\hypersetup{colorlinks=false,pdfborder={0 0 0},bookmarks=false,
  pdftitle={Depth Through Recurrence: Looped Transformers for Flow-Matching TTS},
  pdfauthor={Jiabao Ai, Peng Han, Yuchen Song, Zhengjun Yue}}
\graphicspath{{figs/}}

\title{DEPTH THROUGH RECURRENCE:\\ LOOPED TRANSFORMERS FOR FLOW-MATCHING TTS}

\name{Jiabao Ai\textsuperscript{1,2} \qquad Peng Han\textsuperscript{1,2} \qquad Yuchen Song\textsuperscript{1,2} \qquad Zhengjun Yue\textsuperscript{1,2}}
\address{\textsuperscript{1}Shenzhen Loop Area Institute, Shenzhen, China\\
\textsuperscript{2}The Chinese University of Hong Kong, Shenzhen}

\begin{document}
\ninept
\maketitle

\begin{abstract}
We study how to organize Transformer depth through recurrence in flow-matching text-to-speech, varying the amount, order, and placement of weight reuse. Seven layouts perform 18 block calls per network evaluation under a common training objective and sampler. On Seed-TTS and LibriSpeech-PC, SEQUENCE applies each of nine blocks twice consecutively, retaining competitive intelligibility, speaker similarity, and predicted speech quality at 32 sampling steps with 47.1\% fewer parameters than the unshared baseline. Cycling six blocks three times further reduces model size but raises 32-step word error rates relative to cycling nine blocks twice. At matched parameter counts and executed depth, reuse order and sharing position produce different quality trade-offs. These comparisons depend on sampling budget: Prefix and Suffix have similar 32-step word error rates, but Suffix is worse by 3.44 and 5.97 percentage points at four steps on the two datasets, respectively. Only Middle ranks first or second in mean word error rate at 32 and four steps on both datasets. These results show that the organization of recurrent computation affects synthesis quality, and that reuse layouts should be selected for both the sampling budget and the quality dimensions of interest.
\end{abstract}


\section{Introduction}
\label{sec:intro}
Flow-matching TTS systems such as Voicebox~\cite{le2023voicebox}, E2 TTS~\cite{eskimez2024e2tts} and F5-TTS~\cite{chen2024f5tts} repeatedly evaluate a velocity network during numerical integration~\cite{lipman2023flowmatching}. Computation therefore has two dimensions: network depth and sampling steps. Reusing Transformer blocks over evolving hidden states preserves executed depth with fewer unique parameters. How should this recurrence be organized when the velocity network is itself repeatedly evaluated during sampling?

Cross-layer sharing is established in Universal Transformers~\cite{dehghani2019universal} and ALBERT~\cite{lan2020albert}; Subformer keeps the first and last layers unshared~\cite{reid2021subformer}. Reuse order matters in translation~\cite{takase2023sharing} and language-model pre-training~\cite{gao2026loopie}, while recent looped language models study recurrent computation for efficiency and latent reasoning~\cite{geiping2025recurrentdepth,koishekenov2025etd,hegazy2026grt,wang2026smelt}. In visual generation, ELT adds intra-loop self-distillation~\cite{goyal2026elt}, and Fixed Point Diffusion Models combine implicit recurrence with unshared boundary layers~\cite{bai2024fpdm}. We study how these reuse principles interact with sampling budget and speech-quality criteria in flow-matching TTS.

\begin{figure*}[!t]
\centering
\includegraphics[width=\textwidth]{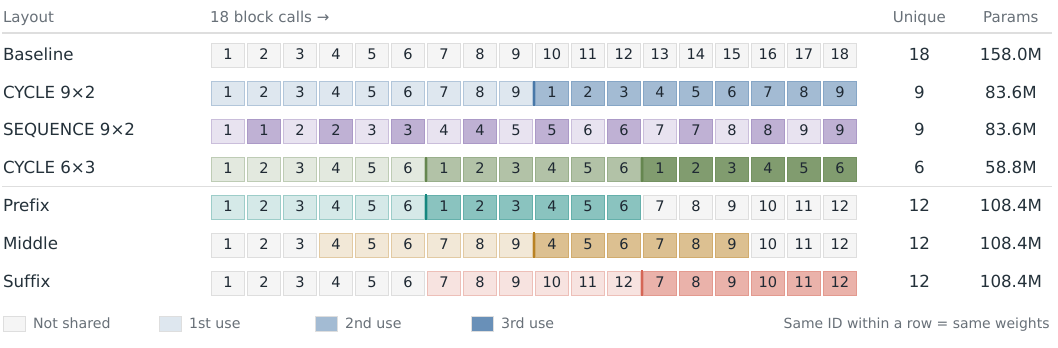}
\caption{Weight-sharing layouts with 18 block calls per network evaluation. Repeated IDs within a row share weights; darker shades indicate later uses. Gray blocks are unshared.}
\label{fig:layouts}
\end{figure*}

We compare seven layouts with 18 block calls per network evaluation under a common objective, sampler, width and training procedure (Fig.~\ref{fig:layouts}). Baseline, CYCLE $9\times2$ and CYCLE $6\times3$ vary sharing amount; CYCLE $9\times2$ and SEQUENCE $9\times2$ compare whole-stack and adjacent-block reuse at matched size; Prefix, Middle and Suffix vary the placement of a repeated six-block segment. Audio samples are available online.\footnote{\normalsize\url{https://jiabaoai67.github.io/loop-f5/}}

On Seed-TTS and LibriSpeech-PC, SEQUENCE retains competitive 32-step quality with 47.1\% fewer parameters than Baseline. Cycling six blocks three times instead of nine twice worsens 32-step intelligibility. Reuse order changes the trade-offs among intelligibility, speaker similarity and predicted quality. Placement interacts with sampling budget: Prefix and Suffix have close 32-step word error rates (WER) but diverge at four steps. Only Middle ranks first or second in mean WER across both budgets and datasets. Its unshared early calls are more removal-sensitive than its tested shared interior. These findings guide layout selection by sampling budget and quality priorities.

\begin{table*}[!t]
\centering\small
\caption{Quality and inference memory at 500k updates. Quality means over four/three inference seeds on Seed-TTS/LibriSpeech-PC; bold marks the best displayed means at each sampling budget. WER: percent. Infer. mem.: peak PyTorch-allocated memory relative to Baseline (H100, fp32, batch 1, 32 steps; vocoder included).}
\label{tab:main}
\begingroup
\setlength{\tabcolsep}{2.6pt}
\renewcommand{\arraystretch}{1.04}
\begin{tabular*}{\textwidth}{@{}c@{\hspace{8pt}}l@{\hspace{8pt}\extracolsep{\fill}}c c r r r r r r@{}}
\toprule
\multicolumn{2}{@{}l}{\textbf{(a) 32 sampling steps}} & \multicolumn{2}{c}{Resources} & \multicolumn{3}{c}{Seed-TTS test-en} & \multicolumn{3}{c}{LibriSpeech-PC} \\
\cmidrule(lr){3-4}\cmidrule(lr){5-7}\cmidrule(l){8-10}
Sharing & Layout & Params (M) & Infer. mem. (\%) & WER$\downarrow$ & SIM-o$\uparrow$ & UTMOS$\uparrow$ & WER$\downarrow$ & SIM-o$\uparrow$ & UTMOS$\uparrow$ \\
\midrule
\textit{None} & Baseline & \makebox[2.5em][r]{158.0} & \makebox[2.5em][r]{100.0} & 2.23 & 0.567 & 3.68 & 2.29 & 0.592 & 3.89 \\
\hline
 & CYCLE 9$\times$2 & \makebox[2.5em][r]{83.6} & \makebox[2.5em][r]{65.5} & 2.60 & 0.562 & 3.90 & 2.14 & 0.600 & 4.06 \\
\textit{Full} & SEQUENCE 9$\times$2 & \makebox[2.5em][r]{83.6} & \makebox[2.5em][r]{65.5} & \textbf{2.03} & \textbf{0.573} & 3.75 & 2.26 & \textbf{0.603} & 3.93 \\
 & CYCLE 6$\times$3 & \makebox[2.5em][r]{58.8} & \makebox[2.5em][r]{54.0} & 2.96 & 0.557 & 3.86 & 2.65 & 0.593 & 4.06 \\
\hline
 & Prefix & \makebox[2.5em][r]{108.4} & \makebox[2.5em][r]{77.0} & 2.36 & 0.563 & 3.88 & 2.24 & 0.569 & \textbf{4.09} \\
\textit{Partial} & Middle & \makebox[2.5em][r]{108.4} & \makebox[2.5em][r]{77.0} & 2.04 & 0.566 & 3.84 & \textbf{2.07} & 0.583 & 4.04 \\
 & Suffix & \makebox[2.5em][r]{108.4} & \makebox[2.5em][r]{77.0} & 2.19 & 0.570 & \textbf{3.91} & 2.10 & 0.593 & 4.02 \\
\midrule
\end{tabular*}
\par\vspace{3pt}
\begin{tabular*}{\textwidth}{@{}c@{\hspace{8pt}}l@{\hspace{8pt}\extracolsep{\fill}}r r r r r r@{}}
\multicolumn{2}{@{}l}{\textbf{(b) 4 sampling steps}} & \multicolumn{3}{c}{Seed-TTS test-en} & \multicolumn{3}{c}{LibriSpeech-PC} \\
\cmidrule(lr){3-5}\cmidrule(l){6-8}
Sharing & Layout & WER$\downarrow$ & SIM-o$\uparrow$ & UTMOS$\uparrow$ & WER$\downarrow$ & SIM-o$\uparrow$ & UTMOS$\uparrow$ \\
\midrule
\textit{None} & Baseline & 10.59 & 0.498 & 2.85 & 17.81 & 0.433 & 2.23 \\
\hline
 & CYCLE 9$\times$2 & 9.37 & 0.496 & 3.08 & 11.36 & 0.444 & 2.59 \\
\textit{Full} & SEQUENCE 9$\times$2 & 9.32 & 0.498 & 3.07 & 11.46 & 0.428 & 2.56 \\
 & CYCLE 6$\times$3 & 11.56 & 0.495 & 3.03 & 16.68 & \textbf{0.456} & 2.80 \\
\hline
 & Prefix & \textbf{6.87} & 0.495 & \textbf{3.24} & 11.20 & 0.415 & \textbf{2.81} \\
\textit{Partial} & Middle & 8.23 & 0.496 & 3.10 & \textbf{10.83} & 0.445 & 2.70 \\
 & Suffix & 10.31 & \textbf{0.499} & 3.07 & 17.17 & 0.423 & 2.41 \\
\bottomrule
\end{tabular*}
\endgroup

\end{table*}

\section{Recurrent Depth for Flow-Matching TTS}
\label{sec:method}
The velocity network $v_\theta(x_t,t,c)$ conditions on noisy speech $x_t$, flow time $t$, and text and reference-speech context $c$. Within one network evaluation, a shared $m$-block segment $F_\theta$ updates its hidden state at visit $k$ as
\begin{equation}
h^{(k+1)}=F_\theta(h^{(k)};t,c).
\end{equation}
Weights, flow time $t$, and conditioning remain fixed across visits; the hidden state is not reset. No visit-index embedding or auxiliary supervision is used.

\textbf{Baseline.} The unshared model has 18 unique blocks and 158.0M parameters.

\textbf{Full recurrence.} CYCLE $9\times2$ executes $1,\ldots,9,\allowbreak 1,\ldots,9$, revisiting the whole stack; SEQUENCE $9\times2$ executes $1,1,\allowbreak 2,2,\allowbreak\ldots,\allowbreak 9,9$, immediately reapplying each block. Both have nine unique blocks and 83.6M parameters. CYCLE $6\times3$ cycles six blocks three times and has 58.8M parameters.

\textbf{Partial recurrence.} A layout has $p$ unshared blocks, an $m$-block segment cycled $K$ times, then $q$ unshared blocks, giving $D=p+mK+q$ executed calls. Prefix, Middle, and Suffix respectively use $(p,m,K,q)=(0,6,2,6)$, $(3,6,2,3)$, and $(6,6,2,0)$. Each has 12 unique blocks, 108.4M parameters, and $D=18$; only the shared segment's position changes.

\textbf{Sampling budget.} Recurrence operates within each network evaluation; sampling repeatedly evaluates the network at successive flow times. With executed depth $D$, $N$ sampling steps, and two classifier-free-guidance (CFG) branches~\cite{ho2022cfg}, the block-call budget is
\begin{equation}
B=2ND.
\label{eq:budget}
\end{equation}
Conditional and unconditional branches count separately, even when batched. All seven layouts retain $D=18$, giving $B=36N$. Changing $N$ varies numerical integration of a fixed velocity network. Block calls quantify computation; latency is measured separately.

\begin{figure*}[!t]
\centering
\includegraphics[width=\textwidth]{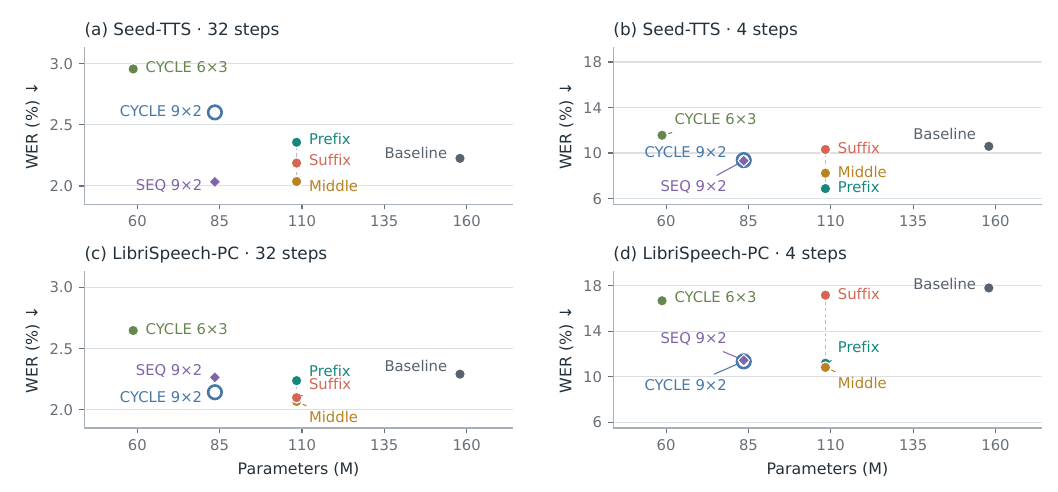}
\caption{WER versus parameter count at 32 and four steps, using the multi-seed means from Table~\ref{tab:main}. Vertical guides connect partial-loop layouts. SEQ: SEQUENCE.}
\label{fig:parameters}
\end{figure*}

\section{Experimental Setup}

\subsection{Training and controls}
All seven layouts use a modified F5-TTS Small implementation~\cite{chen2024f5tts} (width 768) and are trained from scratch on LibriTTS train-clean-100/360 and train-other-500~\cite{zen2019libritts}. The common configuration uses AdamW with peak learning rate $7.5\times10^{-5}$, 20k warm-up updates and linear decay to zero at update 588,931, gradient clipping at 1.0, bf16 mixed precision, up to 307,200 mel frames per update (eight GPUs, accumulation of two), per-rank CFG dropout, and nominal seed 666. Each layout has one training run; comparisons use its 500k-update exponential moving average (EMA) weights.

\subsection{Evaluation}
Seed-TTS test-en~\cite{anastassiou2024seedtts,seedttseval} has 1,088 utterances. For LibriSpeech-PC, we use the F5-TTS subset~\cite{chen2024f5tts,meister2023librispeechpc}: 1,127 utterances from 39 speakers. We use Euler sampling, CFG 2.0, sway $-1$, batch size one and fp32; target durations follow the F5-TTS text-length rule, and Vocos~\cite{siuzdak2023vocos} produces waveforms. We compare 32 and four steps, with intermediate budgets tracing changes in layout performance.

\textbf{WER} measures intelligibility using mean per-utterance Whisper large-v3 error rate~\cite{radford2023whisper}, as in the F5-TTS evaluation code. \textbf{SIM-o} is cosine similarity between prompt and output embeddings from a WavLM-Large/ECAPA-TDNN speaker verifier~\cite{chen2022wavlm}, using 16-kHz audio and per-waveform normalization. \textbf{UTMOS22 strong}~\cite{saeki2022utmos} predicts speech quality; it is not a listening-test score.

\subsection{Uncertainty and resource measurement}
Main-table quality scores average four/three inference seeds on Seed-TTS/LibriSpeech-PC. We first average each utterance over common seeds, then bootstrap paired differences over 666 reference-prompt clusters or 39 speaker clusters, respectively, using 20,000 resamples. The unadjusted 95\% intervals describe evaluation uncertainty conditional on trained checkpoints, not training-seed variance.

Resources are measured on H100 with fp32, batch one and 32-step Seed-TTS synthesis. Peak memory is PyTorch-allocated and includes the vocoder. Sampling real-time factor (RTF) is total sampling time divided by total generated audio duration, excluding vocoding.

\subsection{Diagnostic protocol}
For each call of the 500k EMA models, we measure the relative update $\|\Delta h\|/\|h\|$ on target frames~\cite{jastrzebski2018residual}, and the cosine between two visits' updates to a shared block. We use 128 LibriSpeech-PC utterances and ground-truth/noise interpolants at seven flow times from 0 to 0.9 in both CFG branches (14 settings). For each block--setting pair, statistics are computed per utterance and then averaged. Call removal sets one call's output to its input at every Euler step in both branches~\cite{veit2016residual}, on Seed-TTS with inference seed 666 and 16 steps, without retraining.

\begin{figure*}[!t]
\centering
\includegraphics[width=\textwidth]{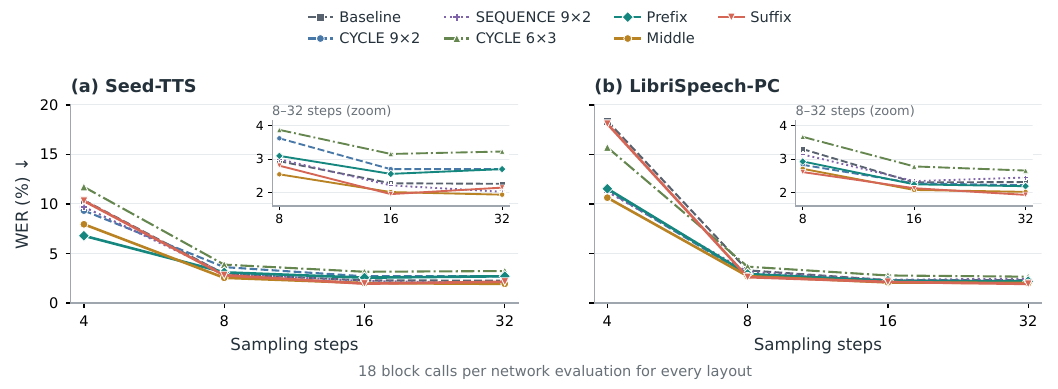}
\caption{WER versus sampling steps for all seven layouts. Curves use one fixed inference seed per dataset (Seed-TTS: 666; LibriSpeech-PC: 0). Insets show 8--32 steps.}
\label{fig:budget}
\end{figure*}

\section{Results and Discussion}
\subsection{Parameter efficiency at matched executed depth}
SEQUENCE uses 83.6M parameters versus Baseline's 158.0M, a 47.1\% reduction, with the same 18 block calls. At 32 steps, its WER is 2.03/2.26\% on Seed-TTS/LibriSpeech-PC versus Baseline's 2.23/2.29\% (Table~\ref{tab:main}). The paired differences are $-0.19$ percentage points (pp), 95\% CI $[-0.48,+0.06]$, and $-0.03$\,pp $[-0.24,+0.16]$. SIM-o and UTMOS means are higher on both datasets, with paired intervals above zero. These comparisons support competitive objective quality at a smaller parameter budget; they do not establish equivalence.

The benefit has limits as sharing increases. CYCLE $9\times2$ has 32-step WER of 2.60/2.14\%; CYCLE $6\times3$ saves another 24.8M parameters but raises WER to 2.96/2.65\% and lowers SIM-o on both datasets. Predicted quality does not decline uniformly: LibriSpeech-PC UTMOS is 4.06 for both. WER is not monotonic in parameter count (Fig.~\ref{fig:parameters}).

CYCLE $9\times2$ and SEQUENCE $9\times2$ reduce peak allocated memory from 877 to 574\,MB (34.5\%); CYCLE $6\times3$ and the partial loops use 474 and 675\,MB. Sampling RTF spans 0.1426--0.1440 across all seven layouts. Parameter storage and peak allocated memory fall, with comparable observed sampling time.

\subsection{Reuse order changes the quality trade-off}
At matched size and depth, SEQUENCE lowers 32-step Seed-TTS WER from CYCLE's 2.60 to 2.03, a paired change of $-0.57$\,pp $[-0.80,-0.35]$, but lowers UTMOS from 3.90 to 3.75. WER rises from 2.14 to 2.26 on LibriSpeech-PC; its paired interval includes zero. SEQUENCE has the highest 32-step SIM-o means on both datasets, but lower UTMOS than CYCLE, with paired intervals below zero. Its Seed-TTS WER is nearly tied with Middle (2.033 versus 2.036). At four steps, CYCLE and SEQUENCE have close WER: 9.37 versus 9.32 on Seed-TTS, and 11.36 versus 11.46 on LibriSpeech-PC. Reuse order changes the quality trade-off without producing a universal winner.

\subsection{Placement interacts with the sampling budget}
At 32 steps, Prefix and Suffix differ by only 0.17/0.14\,pp in WER, with Suffix slightly lower. At four steps, Suffix instead exceeds Prefix by 3.44\,pp, 95\% CI $[2.77,4.14]$, and 5.97\,pp $[3.69,8.70]$. From 32 to four steps, the Suffix-minus-Prefix gap therefore grows by 3.61/6.11\,pp, with both paired intervals above zero. Thus identical model size and executed depth conceal a substantial difference at the most aggressive sampling budget tested.

The intermediate budgets localize this difference (Fig.~\ref{fig:budget}). In the fixed-seed curves, Suffix-minus-Prefix WER changes from $+3.52$/$+6.55$\,pp at four steps to $-0.29$/$-0.32$\,pp at eight. The mean ordering reverses, although the Seed-TTS eight-step interval includes zero. Prefix therefore does not lead throughout the low-step range. The max--min WER spread across the seven layouts shrinks from 4.93/7.74\,pp at four steps to 1.20/0.70\,pp at 16 steps and 1.29/0.73\,pp at 32 steps.

The four-step gap includes more high-error outputs. Across inference seeds, moving from Prefix to Suffix raises the fraction of individual generations with WER above 25\% from 7.54\% to 12.89\% on Seed-TTS and from 12.51\% to 24.82\% on LibriSpeech-PC; 36 of 39 LibriSpeech-PC speakers have higher mean four-step WER. At 32 steps, the corresponding Prefix/Suffix rates are 1.40\%/1.42\% on Seed-TTS and 0.59\%/0.53\% on LibriSpeech-PC.

Quality criteria also change the comparison. At four steps, Prefix exceeds Suffix in UTMOS by 0.164/0.396, with both paired intervals above zero, but has the lowest LibriSpeech-PC SIM-o among all layouts. At 32 steps, Suffix exceeds Prefix in SIM-o by 0.0064/0.0238, with both intervals above zero.

Among partial loops, Middle has the lowest 32-step WER on both datasets and the lowest four-step WER on LibriSpeech-PC. Across all seven layouts, only Middle ranks first or second in mean WER at both budgets on both datasets. Its WER is below Baseline's in all four conditions, with paired intervals below zero except on Seed-TTS at 32 steps. Middle does not lead in SIM-o or UTMOS.

\subsection{Update structure and call sensitivity}
\label{sec:mechanism}
Baseline's first-call mean relative update is 1.6--2.1 across the 14 settings. Removing call 1, 3, 4 or 5 raises WER from 2.28\% to 11.12--102.24\%; removing any call after the fifth changes it by at most 4.64\,pp. Middle's first two calls are unshared and have the largest mean relative updates in every setting; removing either gives 28.84\% or 99.97\% WER, versus 2.02\% unmodified. Removing shared calls 4--14 individually (call 15 untested) changes WER by at most 1.09\,pp. Prefix's first-call mean relative update is at most 0.47, yet removing the second visit of block 4 gives 61.27\% WER. The mean of the last three relative updates is 1.60--2.85 times the median over calls 6--13 for Baseline and Middle, but 0.84--1.35 times for Suffix, which ends on second visits. Among partial loops, only Middle combines large early relative updates with an increase near the output; its boundary blocks are unshared.

Reuse order changes how two visits of a block relate. In SEQUENCE, mean cosines between each block's two visit updates are positive across all 126 block--setting pairs ($+0.42$ to $+0.94$). In CYCLE $9\times2$, block 1 is reapplied to the whole-stack output; its visit updates have negative mean cosine across all 14 settings ($-0.61$ to $-0.33$). Averaged over blocks and settings, the visit cosine is 0.71 for SEQUENCE, 0.35 for CYCLE, and highest among partial loops for Middle (0.62; Prefix 0.47, Suffix 0.52).

\section{Conclusion}
At matched executed depth, our looped flow-matching TTS models retain competitive quality with fewer parameters. Cycling six blocks three times instead of nine twice worsens 32-step intelligibility at 500k updates. Reuse order changes 32-step quality trade-offs but barely affects four-step WER; placement changes four-step WER much more than 32-step WER. Middle ranks first or second in mean WER across all four dataset--budget conditions. Layout choice depends on model size, sampling budget and quality priorities.

The evidence covers one model scale and English objective evaluation. Future work will add listening tests.

\label{sec:endcontent}
\bibliographystyle{IEEEbib}
\bibliography{refs}
\end{document}